\documentclass[10pt, journal]{IEEEtran}
\usepackage[italian, english]{babel}
\usepackage[utf8]{inputenc}
\usepackage{authblk}
\usepackage{graphicx}
\usepackage{amsmath}
\usepackage{caption}
\usepackage{subcaption}
\usepackage{algorithm}
\usepackage {algpseudocode}
\usepackage[justification=centering]{caption}
\usepackage{makecell}
\usepackage{acronym}
\usepackage{multicol}
\usepackage{float}
\usepackage{tabularx,booktabs}
\usepackage{url}
\usepackage{fancyhdr}
\newcolumntype{Y}{>{\centering\arraybackslash}X}
\newcolumntype{?}{!{\vrule width 1.5pt}}
\usepackage{cite}
\usepackage{fancyhdr}
\usepackage[margin=0.680in, top=0.739in, bottom=0.98in]{geometry} 
\usepackage{soul,color}

\newcommand\blfootnote[1]{%
  \begingroup
  \renewcommand\thefootnote{}\footnote{#1}%
  \addtocounter{footnote}{-1}%
  \endgroup
}

\usepackage{titlesec}
\titlespacing\section{0pt}{5pt plus 4pt minus 2pt}{2pt plus 2pt minus 2pt}
\titlespacing\subsection{0pt}{5pt plus 4pt minus 2pt}{2pt plus 2pt minus 2pt}

\acrodef{3GPP}{3rd Generation Partnership Project}
\acrodef{5CN}[5CN]{5G Core Network}
\acrodef{5G-ACIA}{5G Alliance for Connected Industries and Automation}
\acrodef{5G NR}[5G NR]{5G New Radio}
\acrodef{AE}[AE]{Autoencoders}
\acrodef{ANN}[ANN]{Artificial Neural Network}
\acrodef{AGV}[AGV]{Automated Guided Vehicle}
\acrodef{AI}[AI]{Artificial Intelligence}
\acrodef{APN}[APN]{Access Point Name}
\acrodef{1D-CNN}[1D-CNN]{1D Convolutional Neural Network}
\acrodef{CNN}[CNN]{Convolutional Neural Network}
\acrodef{DDNN}[DDNN]{Deep Dense Neural Networks}
\acrodef{DL}[DL]{Deep Learning}
\acrodef{EDF}[EDF]{Earliest Deadline First}
\acrodef{FF}[FF]{Fairness First}
\acrodef{FR1}[FR1]{Frequency Range 1}
\acrodef{FR2}[FR2]{Frequency Range 2}
\acrodef{gNB}[gNB]{gNodeB}
\acrodef{HARQ}[HARQ]{Hybrid Automatic Repeat reQuest}
\acrodef{IIoT}[IIoT]{Industrial Internet of Things}
\acrodef{LR}[LR]{Logistic Regression}
\acrodef{LoS}[LoS]{Line-of-Sight}
\acrodef{ML}[ML]{Machine Learning}
\acrodef{NLoS}[NLoS]{Non Line-of-Sight}
\acrodef{NN}[NN]{Neural Network}
\acrodef{NPN}[NPN]{Non Public Network}
\acrodef{OFDM}[OFDM]{Orthogonal Frequency Division Multiplexing}
\acrodef{PDCCH}[PDCCH]{Physical Downlink Control Channel}
\acrodef{PDSCH}[PDSCH]{Physical Downlink Shared Channel}
\acrodef{PDU}[PDU]{Protocol Data Unit}
\acrodef{PUCCH}[PUCCH]{Physical Uplink Control Channel}
\acrodef{PUSCH}[PUSCH]{Physical Uplink Shared Channel}
\acrodef{RAN}[RAN]{Radio Access Network}
\acrodef{RB}[RB]{Resource Block}
\acrodef{RNN}[RNN]{Recurrent Neural Networks}
\acrodef{RTT}[RTT]{Round-Trip Time}
\acrodef{RUL}[RUL]{Remaining Useful Life}
\acrodef{SCS}[SCS]{Subcarrier Spacing}
\acrodef{SPS}[SPS]{Semi-Persitent Scheduling}
\acrodef{SU}[SU]{Scheduling Unit}
\acrodef{UE}[UE]{User Equipment}
\acrodef{UPF}[UPF]{User Plane Function}
\acrodef{PHY}[PHY]{Physical}
\acrodef{ACK}[ACK]{Acknowledgment}
\acrodef{MCS}[MCS]{Modulation and Coding Scheme}

\acrodef{LSTM}[LSTM]{Long Short Term Memory}
\acrodef{BiLSTM}[BiLSTM]{Bi-directional Long Short Term Memory}
\acrodef{GRU}[GRU]{Gated Recurrent Unit}
\acrodef{E2E}[E2E]{End-to-End}

\title{An End-To-End Analysis of Deep Learning-Based Remaining Useful Life Algorithms for Satefy-Critical 5G-Enabled IIoT Networks}

\author[1,2,*]{Lorenzo Mario Amorosa}
\author[1,2,*]{Nicolò Longhi}
\author[2]{Giampaolo Cuozzo}
\author[1,2]{Weronika Maria Bachan}
\author[3]{Valerio Lieti}
\author[4]{Enrico Buracchini}
\author[1,2]{Roberto Verdone\vspace{-0.4em}}
\affil[1]{DEI, University of Bologna, \textit{Italy}}
\affil[2]{WiLab, CNIT, \textit{Italy}}
\affil[3]{Nokia Solutions and Networks Italia S.p.A., \textit{Italy}}
\affil[4]{Telecom Italia S.p.A., \textit{Italy}}
\affil[*]{\textit {\{lorenzomario.amorosa, nicolo.longhi\}@unibo.it}\vspace{-1cm}}

\begin{document}


\maketitle

\thispagestyle{fancy}
\fancyhf{}
\fancyfoot[C]{\textit{© 2023 IEEE.  Personal use of this material is permitted.  Permission from IEEE must be obtained for all other uses, in any current or future media, including reprinting/republishing this material for advertising or promotional purposes, creating new collective works, for resale or redistribution to servers or lists, or reuse of any copyrighted component of this work in other works.}}
\renewcommand{\headrulewidth}{0pt}






\begin{abstract} 
Remaining Useful Life (RUL) prediction is a critical task that aims to estimate the amount of time until a system fails, where the latter is formed by three main components, that is, the application, communication network, and RUL logic. In this paper, we provide an end-to-end analysis of an entire RUL-based chain. Specifically, we consider a factory floor where Automated Guided Vehicles (AGVs) transport dangerous liquids whose fall may cause injuries to workers. Regarding the communication infrastructure, the AGVs are equipped with 5G User Equipments (UEs) that collect real-time data of their movements and send them to an application server. The RUL logic consists of a Deep Learning (DL)-based pipeline that assesses if there will be liquid falls by analyzing the collected data, and, eventually, sending commands to the AGVs to avoid such a danger. According to this scenario, we performed End-to-End 5G NR-compliant network simulations to study the Round-Trip Time (RTT) as a function of the overall system bandwidth, subcarrier spacing, and modulation order. Then, via real-world experiments, we collect data to train, test and compare 7 DL models and 1 baseline threshold-based algorithm in terms of cost and average advance. Finally, we assess whether or not the RTT provided by four different 5G NR network architectures is compatible with the average advance provided by the best-performing one-Dimensional Convolutional Neural Network (1D-CNN). Numerical results show under which conditions the DL-based approach for RUL estimation matches with the RTT performance provided by different 5G NR network architectures.

\end{abstract}

\begin{IEEEkeywords}
Deep Learning (DL), Remaining Useful Life (RUL), Industrial Internet of Things (IIoT), 5G, NR, Round-Trip Time (RTT), End-to-End (E2E).
\end{IEEEkeywords}



\section{Introduction}

\label{sec:intro}

\ac{RUL} estimation is a predictive task that determine the time until a system fails. Through the utilization of \ac{AI} algorithms, \ac{RUL} prediction provides a means to optimize maintenance strategies, minimize downtime, and foresee system failures, encompassing the integration of sensor data and proactive approaches across diverse industrial sectors \cite{rul4}.\blfootnote{Nicolò Longhi Ph.D. has been funded by Telecom Italia S.p.A.} Generally speaking, a \ac{RUL}-based chain is made of three main components, i.e., (i) the application, (ii) communication network, and (iii) \ac{RUL} logic. To the best of the authors' knowledge, there is no contribution that jointly analyzes all three elements. This paper then aims to fill this gap.

In particular, we consider an \ac{IIoT} safety-critical use case~\cite{3GPP_22.804} where the factory floor contains \acp{AGV} that transport dangerous liquids whose fall may cause injuries to workers. These failure events can be foreseen by \ac{RUL} algorithms. The communication infrastructure is based on \ac{5G NR}, where the AGVs are equipped with 5G \acp{UE} that collect real-time data of their movements and send them to an application server. The \ac{RUL} logic is implemented at the server and consists of a \ac{DL}-based pipeline that analyses the collected data and assesses if there will be liquid falls. Whenever a liquid fall is foreseen, the server sends a command to the \ac{AGV} to avoid this danger. 

For the aforementioned \ac{RUL} scenario, this work provides three main results. First, we performed \ac{E2E} \ac{5G NR}-compliant network simulations to study the \ac{RTT} as a function of the different system parameters, such as the overall system bandwidth, subcarrier spacing, and modulation order. Then, by means of experiments in a real-world industrial plant, we collected \ac{AGV} movement data to train, test and compare 7 \ac{DL} models and a baseline threshold-based algorithm in terms of cost (i.e., a metric related to the number of erroneous predictions) and average advance (a metric quantifying the advance time of the prediction w.r.t the liquid fall). Finally, we evaluate the compatibility between the average advance provided by the best-performing \ac{1D-CNN} and the \ac{RTT} provided by four different \ac{5G NR} network architectures. These architectures align with those foreseen by \ac{3GPP} and \ac{5G-ACIA} \cite{3GPP_23.501, 5GACIA}.

The paper is thus organized as follows. Sec.~\ref{System Model} describes all the features of our \ac{E2E} \ac{5G NR}-compliant network simulator, whereas Sec.~\ref{section: rul} illustrates how we created the dataset and designed the \ac{DL}-based pipeline for \ac{RUL} estimation. Then, Sec.~\ref{RTT Analysis} provides a mathematical formulation of \ac{RTT}, while the corresponding numerical results are presented in Sec.~\ref{numerical results}. Finally, Sec.~\ref{conclusion} summarizes the main findings of our paper.



\section{End-to-End 5G NR-compliant Network Simulator}
\label{System Model}

\subsection{Network Architecture}

\label{subsec: Network Architecture}

\begin{figure}
        \centering
        \begin{subfigure}[b]{0.5\textwidth}
            \centering
            \includegraphics[width=\textwidth]{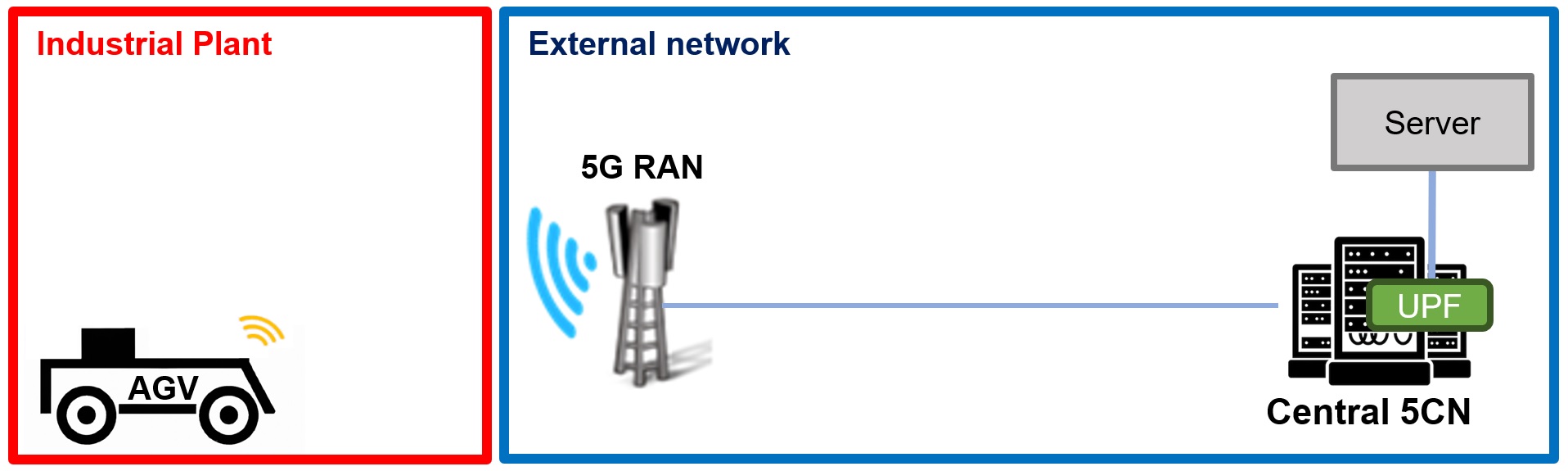}
            \vspace{-5.5mm}
            \caption{{\small Architecture 1.}}
            \vspace{1.5mm}
            \label{fig:architecture_1}
        \end{subfigure}
        \begin{subfigure}[b]{0.497\textwidth}  
            \centering 
            \includegraphics[width=\textwidth]{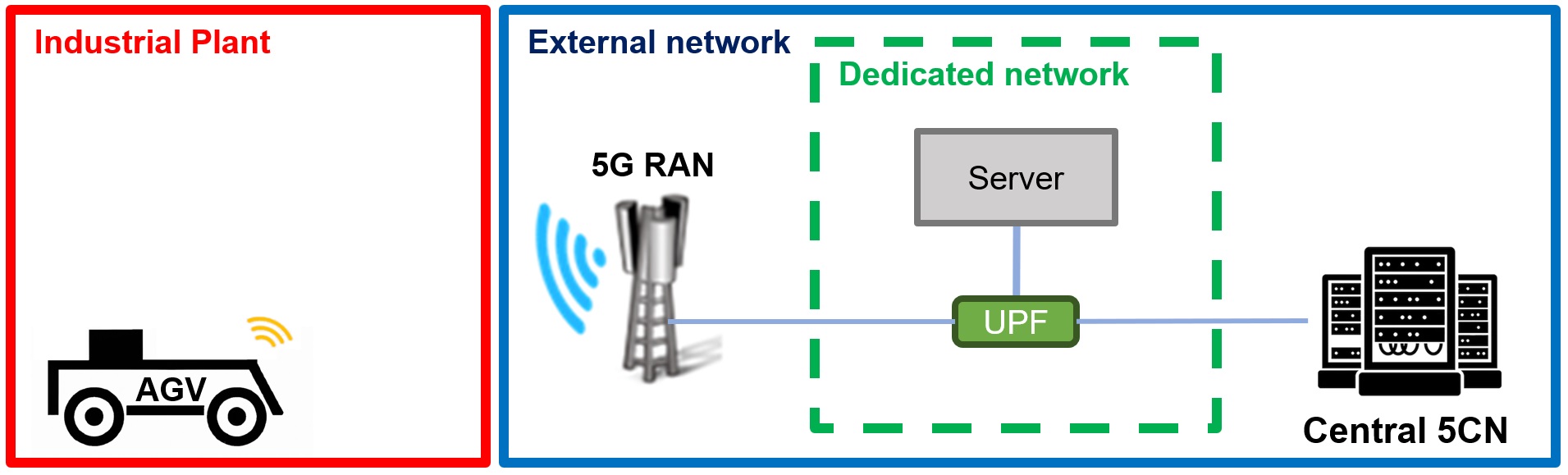}
            \vspace{-5.5mm}
            \caption{{\small Architecture 2.}}    
            \vspace{1.5mm}
            \label{fig:architecture_2}
        \end{subfigure}
        \begin{subfigure}[b]{0.497\textwidth}   
            \centering 
            \includegraphics[width=\textwidth]{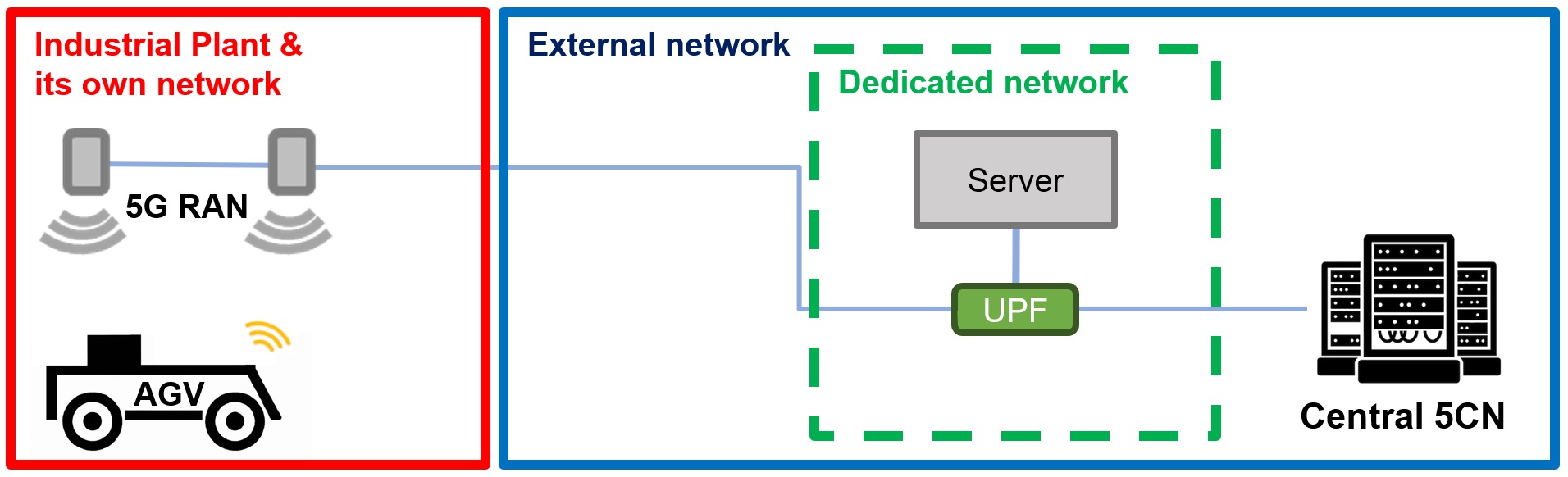}
            \vspace{-5.5mm}
            \caption{{\small Architecture 3.}}
            \vspace{1.5mm}
            \label{fig:architecture_3}
        \end{subfigure}
        \begin{subfigure}[b]{0.497\textwidth}   
            \centering 
            \includegraphics[width=\textwidth]{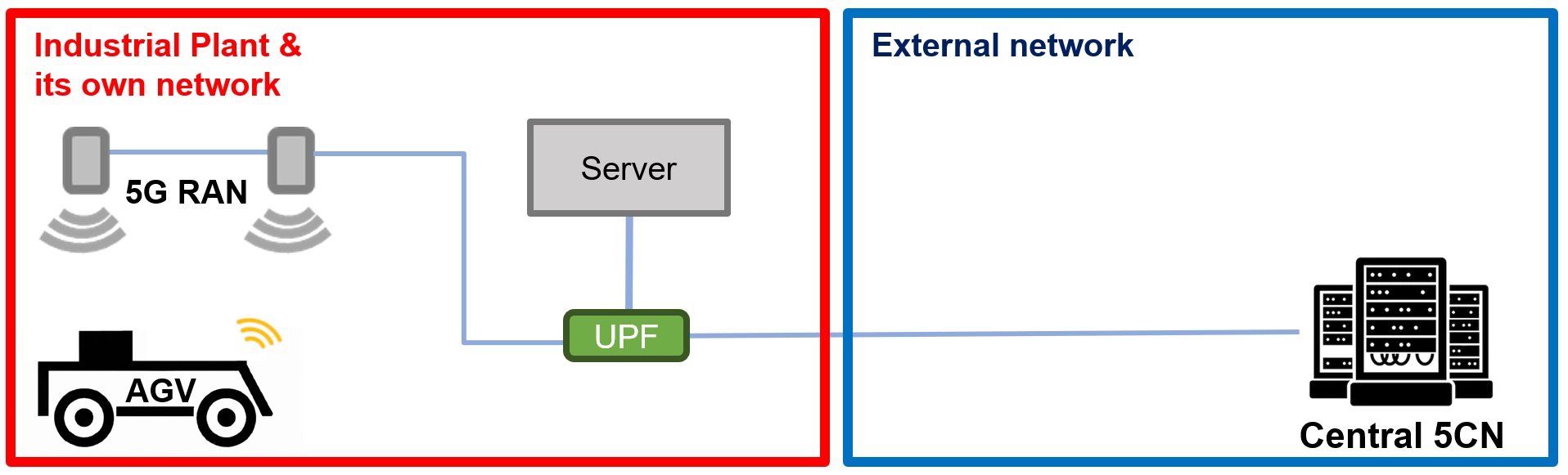}
            \vspace{-5.5mm}
            \caption{{\small Architecture 4.}}
            \vspace{1.5mm}
            \label{fig:architecture_4}
        \end{subfigure}
        \vspace{-6.5mm}
        \caption{Pictorial representation of the considered \ac{5G NR} network architectures.}
        \vspace{-1.4em}
        \label{fig:architectures}
\end{figure}

\thispagestyle{empty}

For the scenario described in Sec.~\ref{sec:intro}, we foresee four different \ac{5G NR} network architectures which are depicted in Fig.~\ref{fig:architectures} and described hereafter. 

\begin{enumerate}
    \item \textbf{Architecture 1}: both the 5G \ac{RAN} and \ac{5CN} are deployed outside the factory (see Fig.~\ref{fig:architecture_1});
    \item \textbf{Architecture 2}: the application server and \ac{UPF} functionalities are hosted at the network operator's premises closer to the industrial plant (see Fig.~\ref{fig:architecture_2}), and 5G \ac{RAN} is deployed outside the factory;
    \item \textbf{Architecture 3}: the 5G \ac{RAN} is deployed in the industrial plant, whereas the application server and \ac{UPF} functionalities are hosted at the public network operator's premises (see Fig.~\ref{fig:architecture_3});
    \item \textbf{Architecture 4}: the 5G \ac{RAN}, the application server, and the \ac{UPF} functionalities are deployed inside the factory, whereas the other \ac{5CN} elements are external (see Fig.~\ref{fig:architecture_4}).
\end{enumerate}

In all cases, the 5G \ac{RAN} is composed of a single \ac{gNB}. These architectures, proposed by TIM, draw inspiration from the \ac{5G-ACIA} documentations \cite{5GACIA}.

\subsection{Traffic model}
\label{subsec: traffic model}

Each \ac{UE} collects information about the \ac{AGV} movements (e.g., positions, axial and angular accelerations, etc.) and sends them to the application server with a fixed periodicity $\tau_{\rm UL}$.

Upon receiving each uplink transmission, the server leverages \ac{DL} to assess whether or not there will be a future fall of the dangerous liquids carried by the considered \ac{AGV}. With a probability $p_{\rm DL}$, the \ac{DL} algorithm will predict a future liquid fall. In this case, the server sends a command to the \ac{UE} to prevent the fall, otherwise, it will not generate any downlink transmission.

\subsection{Channel Model}
The channel model is based on the Gilbert-Elliot model \cite{Hasslinger2008}. It is a 2-state Markov model, where the states are usually referred to as good (G) and bad (B). $g$ and $b$ are the probabilities of correct reception when being in state G and B, respectively, such that $g \gg b$. The transition probability from state G to B is $u$, while the transition probability from state B to G is $v$. Therefore, the reception error rate $p_{\rm{e}}$ in steady state is:
\vspace{-1em}
\begin{align}
p_{\rm e} =& (1-g)\pi_{\rm G} + (1-b)\pi_{\rm B}, \label{error_rate}
\end{align}
where $\pi_{\rm G} = \frac{v}{u+v}$ and $\pi_{\rm B} = \frac{u}{u+v}$ are the probability of being in state G and B, respectively. 




\subsection{Implementation of the 5G NR framework}
In the frequency domain, the available bandwidth $B$ is split in $n_{\rm{RB}}$ \acp{RB}, where each \ac{RB} is composed of 12 \ac{OFDM} subcarriers, such that:

\vspace{-1.4em}
\begin{align}
n_{\rm{RB}} =& \left\lfloor \frac{B}{12\Delta f} \right\rfloor, 
\label{n_RB}
\end{align}
where $\Delta f$ is the \ac{SCS}.

In the time domain, \ac{OFDM} symbols are grouped into slots, in particular, 14 \ac{OFDM} symbols form one slot. However, in \ac{5G NR}, it is also possible to have communications over fractions of slots, the so-called ``mini-slots''. In this regard, we used mini-slots composed of 7 \ac{OFDM} symbols each. 

We consider the same uplink messages defined in \cite{Cuozzo2022}, with the inclusion of the downlink ones, as summarized in the following:
\begin{itemize}
    \item \ac{PUCCH}, used by \acp{UE} to ask the \ac{gNB} for being scheduled. It occupies 1 \ac{RB} and 7 \ac{OFDM} symbols.
    \item \ac{PDCCH}, used by the \ac{gNB} to inform the \acp{UE} about the uplink/downlink scheduling outcome. It occupies 1 \ac{RB} and 7 \ac{OFDM} symbols.
    \item \ac{PUSCH}, used by \acp{UE} to transmit \ac{PHY} \acp{PDU}. It occupies 1 or more \ac{RB} and 4 \ac{OFDM} symbols depending on the network load, scheduling policy, etc.
    \item \ac{PDSCH}, used by the \ac{gNB} to transmit \ac{PHY} \acp{PDU} to \acp{UE}. It occupies 1 or more \ac{RB} and 4 \ac{OFDM} symbols.
    \item \ac{HARQ} \ac{ACK}, used to inform the sender about the outcome of the uplink/downlink transmission. It occupies 1 \ac{RB} and 2 \ac{OFDM} symbols.
\end{itemize}
It is worth mentioning that a \ac{PUSCH}/\ac{PDSCH} transmission is followed by the corresponding \ac{HARQ} \ac{ACK} and the overall process has a duration of one mini-slot. Indeed, we assume half-duplex communications, and that one \ac{OFDM} symbol is needed to switch from transmission to reception and vice-versa.

\subsection{Implementation of the 5G NR dynamic scheduling}
\label{subsection:5G scheduler}
The \ac{gNB} makes scheduling decisions, i.e., assigns \acp{RB} and \ac{OFDM} symbols to \acp{UE}, both in uplink and downlink, every $T_{\rm{SRP}}$. Specifically, $T_{\rm{SRP}}$ is formed by 8 mini-slots, where half are dedicated to the control plane (i.e., \ac{PUCCH}, \ac{PDCCH} and \ac{HARQ} \acp{ACK}) and the other half to the data plane (i.e., \ac{PUSCH}, and \ac{PDSCH}). 

The control plane resource assignment is fixed a-priori, i.e., each \ac{UE} knows when to transmit/receive \acp{PUCCH}/\acp{PDCCH}, respectively, according to a predetermined pattern that repeats over time and depends on the number of resources available, as well as network load. On the opposite, the data plane resources are scheduled based on the current traffic needs, according to the \ac{5G NR} scheduling mechanisms denoted as \textit{dynamic scheduling}. In particular, \acp{UE} willing to transmit data in the uplink have to first send a \ac{PUCCH} before receiving the indications via \ac{PDCCH} on how to transmit the \acp{PUSCH}. Additionally, in the downlink, the \ac{gNB} exploits the \acp{PDCCH} to tell the \acp{UE} when (and how) they will receive data. To make the scheduling decisions, the \ac{gNB} exploits the two scheduling policies defined in \cite{Cuozzo2022}, by applying them independently to the uplink and downlink traffic flows. 




\section{DL-based RUL estimation pipeline}
\label{section: rul}

\thispagestyle{empty}

In this section, we describe (i) how we created a dataset from experiments in a real-world industry plant, and (ii) the \ac{DL}-based pipeline for \ac{RUL} estimation. It is worth mentioning that some details are omitted due to confidentiality reasons.

\subsection{Data collection}
\label{sec:rul_estimation_tasks}

We performed an experimental campaign where we collected real-time data (accelerations, positions, etc.) of the \ac{AGV} moving within the industrial pilot line of BIREX\footnote{BIREX is an Italian Competence Center for Industry 4.0 (see https://bi-rex.it/)}. Several sessions were registered, where, during each session, the \ac{AGV} carried a bottle of water and was forced, at some point in time, to perform a sudden change in its path, thereby leading to the fall of the bottle. A custom script registered the fall event's timestamp to correctly label the closest sensor's data as a \textit{Fault} event and, consequently, all the others were labeled as \textit{Non-Fault}. The data collected within each session constitute time series data, as they record the \ac{AGV} movement over time with regular intervals and timestamps, resulting in a sequence of ordered observations. In this context, we formulate the \ac{RUL} prediction problem as a binary classification task, where the objective is to predict whether a \textit{Fault} event occurs within a certain margin $m$. This margin defines the classification task since it determines the number $m$ of time series samples prior to a \textit{Fault} event that can be labeled as \textit{Fault} and should trigger the server to send a command to the \ac{AGV} for avoiding potential failures.

\subsection{Data pre-processing}
\label{sec:pre-processing}
The real-time data collected by the \ac{AGV} has been manipulated via different pre-processing steps that are listed hereafter:

\begin{itemize}
    \item \textit{Class weighting} \cite{rul3} to cope with the skewed distribution where the majority of the data points correspond to the normal operating conditions and a minority of the data points correspond to the failed state, as usual in \ac{RUL} scenarios; 
    \item \textit{Feature creation} to obtain a more suitable representation of the physical phenomenon. Specifically, we created the mean, maximum, minimum, and standard deviation over multiple fixed-length windows \cite{brownlee}, as well as the derivative of those time series as the difference between subsequent data points;  
    \item \textit{Differencing} to remove seasonality from the time series data \cite{stationarity}. This was done by subtracting from each data point the mean acceleration value for its position and orientation;
    \item \textit{Standardization} to ensure that all features are on a similar scale and improve the performance and convergence of \ac{DL} models during training.
\end{itemize}

\begin{figure*}[!ht]
\begin{minipage}[b]{0.48\linewidth}
\centering
\includegraphics[width=\linewidth]{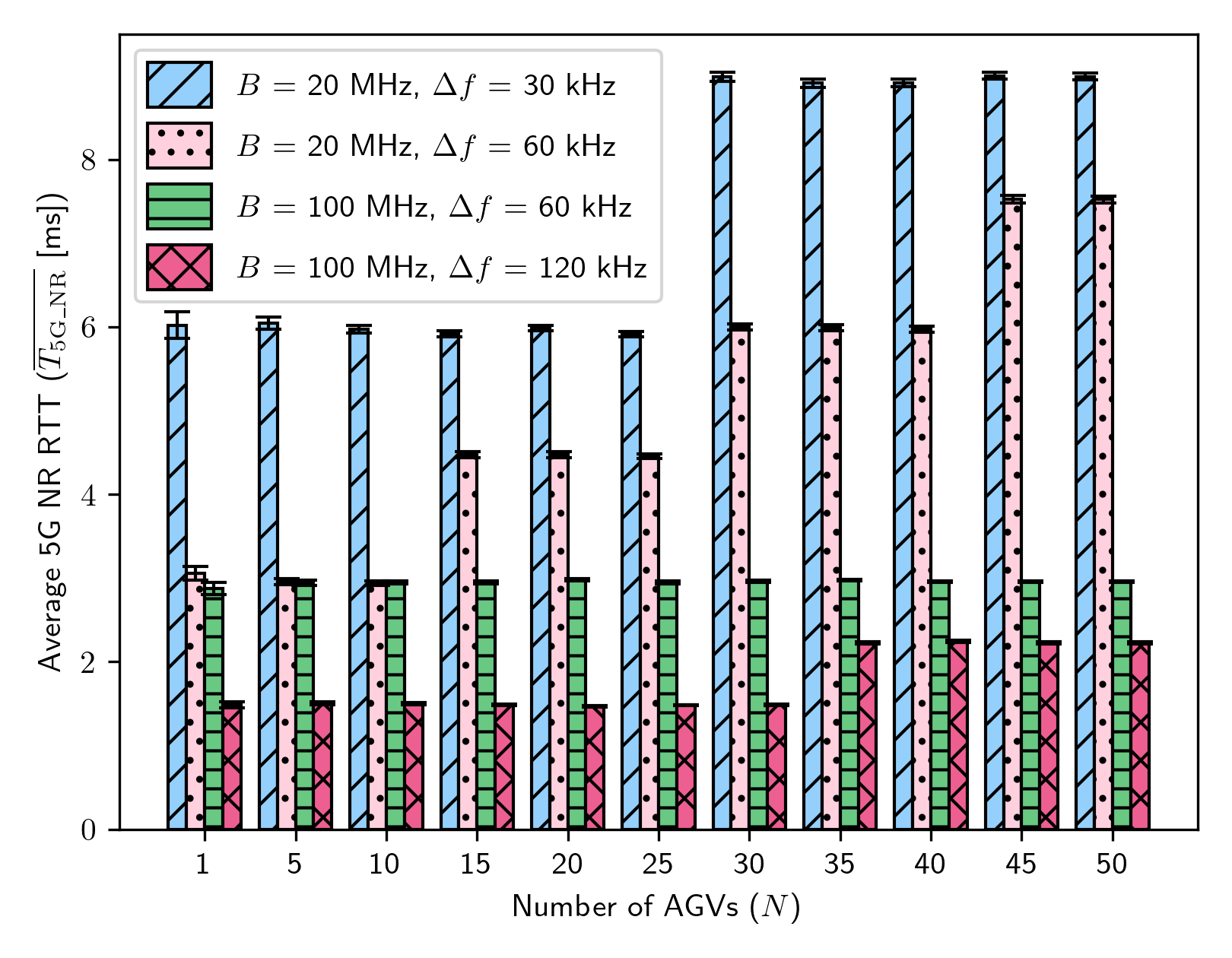}
\vspace{-7mm}
\caption{$\overline{T_{\rm{5G\_NR}}}$ as a function of $N$, $B$ and $\Delta f$. We set $M = 256$ and $T_{\rm{CN}} = 0$ ms.\vspace{-1.5em}}
\label{fig:plot_256QAM}
\end{minipage}
\hspace{0.5cm}
\begin{minipage}[b]{0.48\linewidth}
\centering
\includegraphics[width=\linewidth]{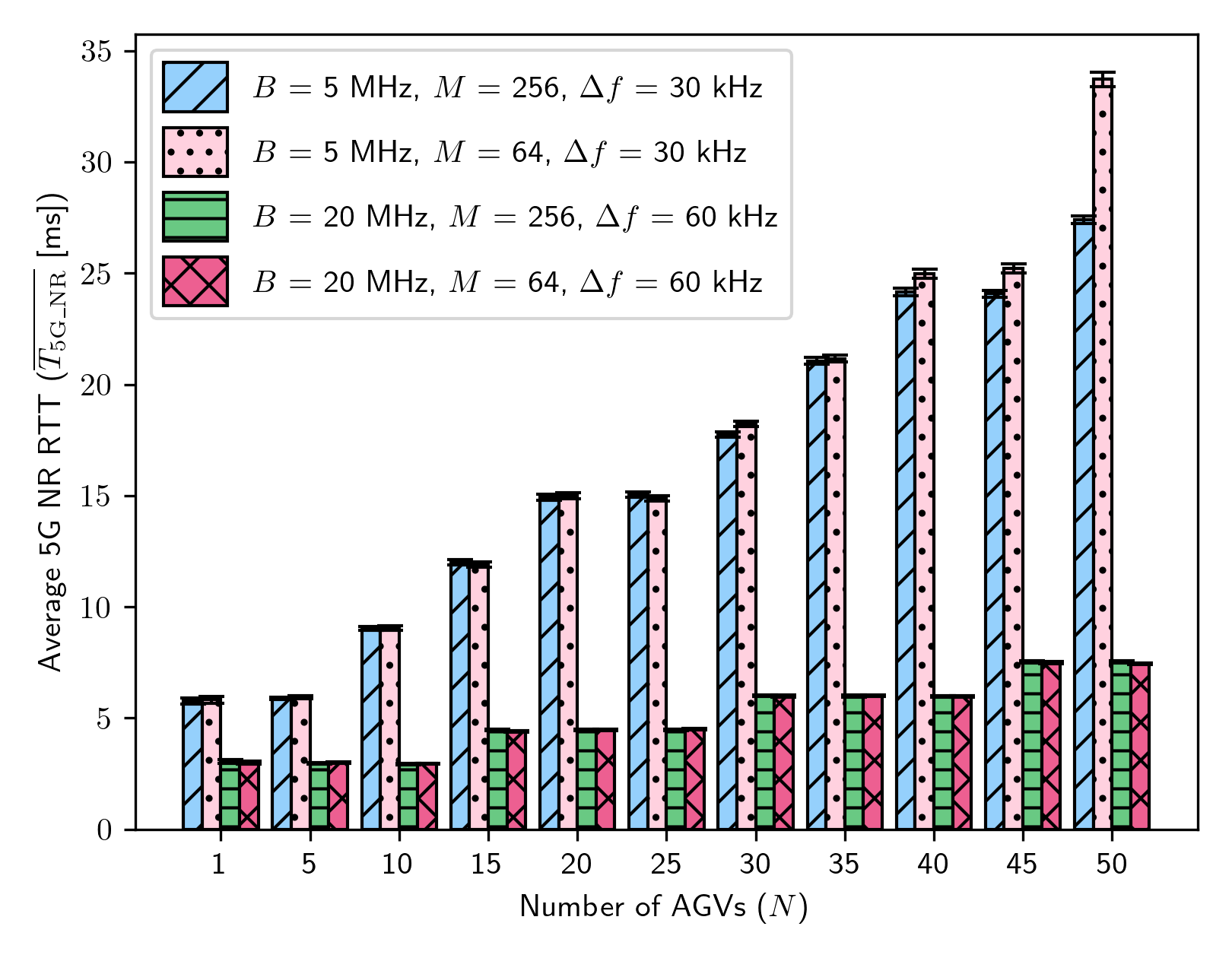}
\vspace{-7mm}
\caption{$\overline{T_{\rm{5G\_NR}}}$ as a function of $N$, $B$, $\Delta f$, and $M$. We set $T_{\rm{CN}} = 0$ ms.\vspace{-1.5em}} 
\label{fig:plot_64QAMvs256QAM}
\end{minipage}
\end{figure*}

\subsection{DL-based pipeline}
\label{sec:ai_pipeline}
To perform \ac{RUL} estimations, the server implements a \ac{DL}-based pipeline that is formed by a \ac{DL} model and a threshold-based algorithm. Indeed, the former is trained on the data collected during the hands-on measurement campaign to predict the \ac{RUL}, and the latter transforms the \ac{DL} output to either zero or one as the \ac{RUL} task is tackled as a binary classification problem (see Sec.~\ref{sec:rul_estimation_tasks}). 

The \ac{DL}-based pipeline was trained, optimized, and tested by exploiting the manipulated dataset described in Sec.~\ref{sec:pre-processing}. Specifically, the dataset was further partitioned into 4-folds, i.e., 1) a training set to train the \ac{DL} model, 2) a validation set to monitor the \ac{DL} model performance at training time, 3) another validation set to define the optimal value of the threshold, and 4) a test set to evaluate the \ac{DL}-based pipeline performance.

In particular, the validation set of step 3) is used iteratively in a procedure that aims to find the optimal threshold which minimizes the cost function $C$, whose expression for a \ac{DL} model $X$ over a set of $K$ time series $S = \{S_1, S_2, \ldots, S_{\rm K}\}$ is as follows:

\vspace{-1.5em}
\begin{align}
C = \sum_{k=1}^{K}\sum_{p=1}^{P_{\rm k}}C_{\rm FP} + \sum_{k=1}^{K}\sum_{q=1}^{Q_{\rm k}}C_{\rm FN}(s_{\rm q}, S_{\rm k}, m),
\label{eq:cost_model}
\end{align}

where $P_{\rm k}$ is the number of false positive samples for the $k$-th time series, $Q_{\rm k}$ is the number of false negative samples for the $k$-th time series, $C_{\rm FP}$ is the cost for a false positive sample, $C_{\rm FN}$ is the cost for a false negative sample. The expressions for $C_{\rm FP}$ and $C_{\rm FN}$ are the following:

\vspace{-1em}
\begin{align}\label{eq:false_negatives_positives_cost}
\begin{split}
  C_{\rm FP} &= 0.2 \\
  C_{\rm FN}(s_{\rm q}, S_{\rm k}, m) &= m - |S_{\rm k}| + q,
\end{split}
\end{align}

where $m$ is the margin, $S_{\rm k}$ is a time series, and $q$ is the index of $s_{\rm q}$ in $S_{\rm k}$. We set the cost of false positives $C_{\rm FP}$ constant, regardless of their occurrence in the time series or the margin, while the cost of false negatives $C_{\rm FN}$ increases the closer the sample is to the anomalous event. This design choice is inherently bounded to safety-critical applications, where false negative samples are considered the major risk to deal with.

However, the most important metric to assess the quality of the \ac{DL}-based pipeline is the average advance function $\bar{a}({D}_{\rm X})$, where ${D}_{\rm X} = \{{s}_1, {s}_2, \ldots, {s}_{\rm K}\}$ is the set of the first samples in the time series detected correctly as faulty by a given model $X$ for a given margin $m$. It indicates the amount of time before the actual fault occurs and it is defined as follows:

\vspace{-1em}
\begin{align}
\bar{a}({D}_{\rm X}) =\frac{\sum_{i=1}^{K}a({s}_{\rm i})}{K},
\label{eq:mean_anticipation}
\end{align}

where $a({s}_{\rm i})$ is the advance function which indicates the amount of time before the actual fault occurs after sample ${s}_{\rm i}$.



\vspace{-0.8em}
\section{Round-Trip Time Analysis}
\label{RTT Analysis}
We define the \ac{RTT} for a single \ac{UE} as the time from which the movements' data are generated at the application layer to the time it receives the corresponding command. In particular, its expression is given by:

\thispagestyle{empty}

\vspace{-1.5em}
\begin{align}
R = & T_{\rm {P\_S}} + 2 \cdot (T_{\rm{P\_gNB}} + T_{\rm{P\_UE}} + T_{\rm{CN}}) + T_{\rm{RAN\_UL}} + \nonumber \\
& + T_{\rm{RAN\_DL}} + T_{\rm{A}} =  \nonumber \\
&= T_{\rm{P\_S}} + T_{\rm{5G\_NR}} + T_{\rm{A}}, 
\label{RTT_expression} 
\end{align}

\vspace{-0.5em}
where: 
\begin{itemize}
    \item $T_{\rm{P\_S}}$ is the processing time of the \ac{DL}-based pipeline described in Sec.~\ref{section: rul};
    \item $T_{\rm{P\_gNB}}$ is the time needed by the \ac{gNB} to process a \ac{PHY} \ac{PDU}, i.e., to eliminate all headers along the 5G protocol stack;
    \item $T_{\rm{P\_UE}}$ is the time needed by the \ac{UE} to process a \ac{PHY} \ac{PDU};
    \item $T_{\rm{CN}}$ is the delay introduced by the \ac{5CN};
    \item $T_{\rm{RAN\_UL}}$ is the time needed by the \ac{UE} to successfully perform an uplink transmission;
    \item $T_{\rm{RAN\_DL}}$ is the time needed by the \ac{gNB} to successfully perform a downlink transmission;
    \item $T_{\rm{A}}$ is the time needed to execute the actuation command;
    \item $T_{\rm{5G\_NR}}$ is the overall delay contribution caused by \ac{5G NR}.
\end{itemize}

We then introduce the term $\overline{T_{\rm{5G\_NR}}}$, as well as the average \ac{RTT} $\bar{R}$, both averaged among the total number of 5G \acp{UE}, i.e., $N$, and the total number of simulations $N_{\rm S}$.

\section{Numerical results}
\label{numerical results}

\begin{table}[!t]
\centering
\begin{tabular}{|c|c|c|} 
\hline
\textbf{Parameter} & \textbf{Description} & \textbf{Value} \\
\hline
$B$ & Overall system bandwidth & \{5, 20, 100\} MHz\\
$\Delta f$ & Subcarrier spacing & \{30, 60, 120\} kHz \\
$M$ & Modulation order of & \{64, 256\} \\
& PUSCHs/PDSCHs transmissions & \\
$p_{\rm{e}}$ & Reception error rate in steady state & $1\%$ \\
$P_{\rm{UL}}$ & Uplink payload & 32 B\\
$P_{\rm{DL}}$ & Downlink payload & 1 B\\
$p_{\rm DL}$ & Downlink generation probability & $10\%$ \\
$T_{\rm{P\_gNB}}$ & \ac{gNB} processing time & 7 \ac{OFDM} symbols\\
$T_{\rm{P\_UE}}$ & \ac{UE} processing time & 7 \ac{OFDM} symbols\\
$T_{\rm{CN}}$ & Delay introduced by the \ac{5CN} & \{1, 2, 7\} ms \\
$T_{\rm{SRP}}$ & Scheduling periodicity & 8 mini-slots\\
$T_{\rm{S}}$ & Simulation time & 10 s \\
$\tau_{\rm{UL}}$ & Uplink periodicity & 100 ms\\
$H$ & 5G protocol stack header & 72 B\\
$N_{\rm S}$ & Number of simulations & 20\\
\hline
\end{tabular}
\vspace{-1.5mm}
\caption{Simulation parameters \vspace{-1.5em}}
    \label{tab:RAN_parameters}
\end{table}

\begin{table*}[t!]
  \centering
\begin{tabularx}{\textwidth}{ |c?Y|Y?Y|Y?Y|Y? }
 \cline{2-7}
   \multicolumn{1}{c?}{} &
   \multicolumn{2}{c?}{$m = 5$} & \multicolumn{2}{c?}{$m = 10$} & \multicolumn{2}{c?}{$m = 15$} \\
 \hline
Model $X$ & $C$ & $\bar{a}({D}_{\rm X})$ & $C$ & $\bar{a}({D}_{\rm X})$ & $C$ & $\bar{a}({D}_{\rm X})$ \\
 \Xhline{4\arrayrulewidth} 
 BASELINE & 80.80 & 0.24s & 306.40 & 0.46s & 448.20 & 1.07s \\
 \Xhline{4\arrayrulewidth} 
 LR & 96.00 & 0.32s & 221.20 & 0.44s & 679.60 & 0.45s \\
 \hline
 DDNN & 43.40 & 0.27s & 142.00 & 0.66s & 311.20 & 0.95s \\
 \hline
 \textbf{1D-CNN} & \textbf{28.80} & \textbf{0.27s} & \textbf{114.40} & \textbf{0.80s} & \textbf{197.60} & \textbf{1.33s} \\
 \hline
 AE & 2396.40 & 0.39s & 2666.80 & 0.90s & 2561.80 & 1.40s \\
 \hline
 LSTM & 95.40 & 0.20s & 346.40 & 0.34s & 569.80 & 0.81s \\
 \hline
 BiLSTM & 61.60 & 0.28s & 272.40 & 0.48s & 689.40 & 1.08s \\
 \hline
 GRU & 85.40 & 0.23s & 290.80 & 0.68s & 539.80 & 0.92s \\
\hline
\end{tabularx}
\vspace{-1.5mm}
\caption{Cost and average advance function of seven \ac{DL} models and a baseline threshold-based approach for three different margins.\vspace{-1.5em}}
  \label{table_model}
\end{table*}

\subsection{Performance of the 5G NR network}
\label{sec:5g_nr_performance}

Simulations parameters, if not otherwise specified, are reported in Table~\ref{tab:RAN_parameters}. However, it is important to underline that:

\begin{itemize}
    \item We assume that $p_{\rm{e}} = 1\%$ and $M = \{64, 256\}$ for \ac{PUSCH} and \ac{PDSCH} communications, but \ac{PUCCH}, \ac{PDCCH} and \ac{HARQ} \ac{ACK} receptions are error-free as they are transmitted with the most conservative modulation order, i.e., $p_{\rm{e}} = 0 \%$ due to $M=4$; 
    \item {We set $p_{\rm DL} = 10 \%$ to have, on average, multiple \ac{PDSCH} transmissions per \ac{UE} within the simulation time $T_{\rm{S}}$.}
    \item Among the four mini-slots per $T_{\rm{SRP}}$ dedicated to data plane resources (see Sec.~\ref{subsection:5G scheduler}), three of them are dedicated to \acp{PUSCH}, and only one to \acp{PDSCH}, due to the uplink-oriented nature of the considered traffic model (see Sec.~\ref{subsec: traffic model});
    \item All results will show a confidence interval with a probability of $90\%$.
\end{itemize}

We start the analysis by investigating the impact of different network parameters on the term $T_{\rm{5G\_NR}}$ of Eq.~\eqref{RTT_expression}, where we set $T_{\rm CN} = 0$ ms to be independent of the four architectures defined in Sec.~\ref{subsec: Network Architecture}. 

In particular, Fig.~\ref{fig:plot_256QAM} shows $\overline{T_{\rm{5G\_NR}}}$ as a function of $N$, $B$, $\Delta f$, and by setting $M = 256$. It can be observed that a wider bandwidth provides better performance due to a higher number of \acp{RB} (see Eq.~\eqref{n_RB}), for both the control and data plane. Quite interestingly, there is also a benefit when $\Delta f$ increases for a fixed value of $B$. This is because the shorter $T_{\rm{SRP}}$ duration translates into more transmission opportunities per unit of time. As expected, $\bar{R}$ increases with $N$ but exhibits a stepwise behavior. The reason is that, for sufficiently high values of $N$, it is not possible to serve all \acp{UE} within one $T_{\rm{SRP}}$ because of a shortage of \textit{control plane} resources; therefore, some \acp{UE} have to transmit/receive their \acp{PUCCH}/\acp{PDCCH} in the next scheduling periodicity with a consequent non-negligible increase in the average \ac{RTT}.

To assess the impact of different modulation orders for a different set of bandwidths, Fig.~\ref{fig:plot_64QAMvs256QAM} depicts $\overline{T_{\rm{5G\_NR}}}$ as a function of $N$, $B$, $\Delta f$, and by considering $M = 64$ and $M = 256$. It can be clearly noted that, regardless of the values of $N$, both modulation orders provide the same performance when $B = 20$ MHz. This is no longer true in the case of $N \ge 30$ and $B = 5$ MHz, where it is preferable to have a higher modulation order, i.e., $M = 256$. This is because, otherwise, there is a shortage of \textit{data plane} resources, and some \acp{UE} are forced to transmit/receive their \acp{PUSCH}/\acp{PDSCH} in the next scheduling periodicity.



\subsection{Performance of the DL-based RUL estimation pipeline}
\label{sec:ai_pipeline_performance}

\thispagestyle{empty}

In this section, we present the performance of the \ac{DL}-based \ac{RUL} estimation pipeline that has been constructed, trained, and tested as described in Sec.~\ref{section: rul}. 

Specifically, Table~\ref{table_model} illustrates the cost $C$ and average advance function $\bar{a}({D}_{{\rm X}})$ (see Eqs.~\eqref{eq:cost_model} and \eqref{eq:mean_anticipation}) provided by seven different \ac{DL} models, including \ac{LR}, \ac{DDNN}, \ac{AE}, \acf{1D-CNN}, \ac{LSTM}, \ac{BiLSTM}, \ac{GRU} \cite{cnn, AE, lstm}, and a baseline threshold-based approach which works over the raw data. Three margins have been considered, that is, $m = \{5, 10, 15\}$. It can be clearly seen that a trade-off exists for all the considered models, as low margins (i.e., $m = 5$) correspond to low average advance and cost, while high margins (i.e., $m = 15$) correspond to higher average advance and cost.  

However, \ac{1D-CNN} is the best-performing model when considering both metrics because it better captures the local temporal patterns present in the data. More complex memory-based models, such as \ac{LSTM}, \ac{BiLSTM} and \ac{GRU}, are not effective in this particular \ac{RUL} estimation task. This is because only a few input samples are relevant for predicting the liquid fall, whereas recurrent models are designed to capture long-term dependencies and patterns in time series data. Another notable observation is that the reconstruction error, which autoencoders seek to minimize, may not be an effective indicator for predicting the \ac{RUL} because these models exhibited markedly inferior performance compared to the others.

\subsection{Performance of the entire RUL chain}

\begin{figure*}[ht!]
\centering
\includegraphics[width=0.97\linewidth]{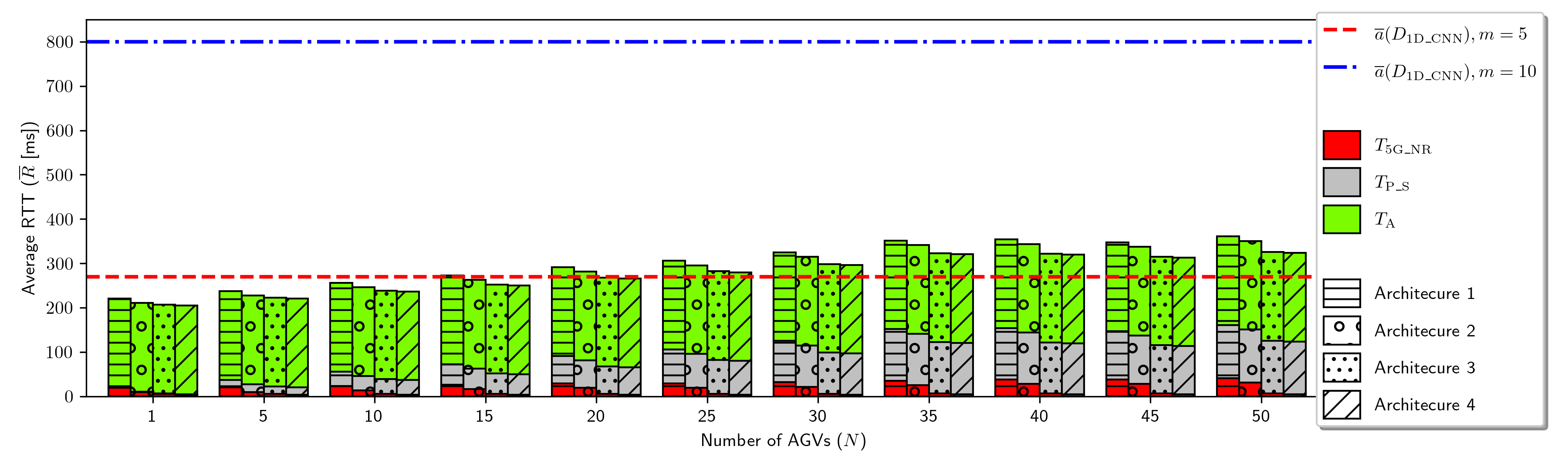}
\vspace{-3mm}
\caption{\label{fig:Total_RTT} Average \ac{RTT} $\bar{R}$, as a function of $N$, and the four network architectures described in Sec.~\ref{subsec: Network Architecture}, when considering \ac{1D-CNN} as \ac{DL} model, and two average advance values $\bar{a}({D}_{\rm 1D\_CNN})$ for $m = 5$ and $m = 10$. \vspace{-1.5em}}
\end{figure*}

In this section, we finally assess whether or not the average \ac{RTT} provided by the four \ac{5G NR} \ac{IIoT} architectures described in Sec.~\ref{subsec: Network Architecture} is sufficiently lower than the average advance provided by \ac{1D-CNN}, i.e., the best performing \ac{DL} model according to the results presented in Sec.~\ref{sec:ai_pipeline_performance}. To this aim, we exploited the \ac{5G NR} network simulator described in Sec.~\ref{System Model} by considering the same settings described in Sec.~\ref{sec:5g_nr_performance}. However, differently from Sec.~\ref{sec:5g_nr_performance}, this \ac{RTT} analysis also considers the first and third terms of Eq.~\eqref{RTT_expression}, as well as the four network architectures described in Sec.~\ref{subsec: Network Architecture}. In particular:

\begin{itemize}
    \item $T_{\rm{P\_S}}$ ranges from $1.2$ ms to $119.2$ ms, depending on the number of  \acp{UE}. This range derives from experimental tests that consider \ac{1D-CNN} processing when considering an i9-11900K processor with 128 GB of RAM and up to 50 parallel flows\footnote{Additional tests were made using i5-6200U processor and 16 GB of RAM, but they led to unsuitable performance (i.e., processing time above 1 second for 30 \acp{AGV}).};
    \item $T_{\rm A} = 200$ ms and is taken from a commercial product.\footnote{See: https://www.hitbotrobot.com/product/z-efg-12-robotic-gripper/}
    \item Due to their structure, and by leveraging real-world data coming from the TIM infrastructure, we consider that Architecture 1 and 2 work with $B = 5$ MHz, $\Delta f = 30$ kHz, $M = 256$, and are characterized by $T_{\rm{CN}} = 7$ ms, and $T_{\rm{CN}} = 2$ ms, respectively. Conversely, Architecture 3 and 4 operate with $B = 100$ MHz, $\Delta f = 120$ kHz, $M = 64$, and are characterized by $T_{\rm{CN}} = 2$ ms, and $T_{\rm{CN}} = 1$ ms, respectively.
\end{itemize}
 
As a result, Fig.~\ref{fig:Total_RTT} shows the average \ac{RTT}, $\bar{R}$, as a function of $N$, and the four network architectures described in Sec.~\ref{subsec: Network Architecture}. Average advance values $\bar{a}({D}_{\rm 1D\_CNN})$ for $m = 5$ and $m = 10$ are represented with dashed horizontal lines. The different colors of each bar represent a diverse delay contribution, that is, $T_{\rm 5G\_NR}$, $T_{\rm P\_S}$, and $T_{\rm A}$ (see Eq.~\eqref{RTT_expression}), while different architectures are represented by different bar patterns.

Notably, when $N \leq 10$, all architectures provide an $\bar{R}$ lower than the average advance margin with $m = 5$. As expected, Architecture 3 and 4 are the best-performing ones, as they are characterized by higher bandwidths, \ac{SCS}, and smaller core network delays; indeed, they still yield sufficiently lower average \acp{RTT} for $N = 15$ and $N = 20$. Nevertheless, when $N \geq 25$, it is necessary to change the \ac{RUL} task by increasing $m$ up to 10, independently of the considered network architectures. However, as described in Sec.~\ref{sec:ai_pipeline_performance}, a higher margin leads to an increase in the cost $C$, i.e., the number of erroneous fall predictions, which could compromise the safety conditions within the factory. Finally, it is worth highlighting that the average \ac{RTT} is mainly affected by application-specific parameters, i.e., $T_{\rm P\_S}$ and $T_{\rm A}$, and optimizations of \ac{5G NR} are not really needed for the considered \ac{IIoT} scenario and settings. 


\section{Conclusions}
\label{conclusion}
In this paper, we performed an \ac{E2E} analysis of a \ac{RUL} estimation problem, where all its three main elements, namely application, communication system, and \ac{RUL} logic, have been studied. In particular, we considered a safety-critical \ac{IIoT} application where \ac{5G NR} is used to collect movements data from \acp{AGV} carrying dangerous liquids, while \ac{DL} algorithms are trained to foresee the potential liquid falls.

\thispagestyle{empty}

The problem is analyzed by considering 4 different \ac{5G NR} architectures, 7 \ac{DL}-based and 1 threshold-based algorithm, as a function of different system parameters. The main findings of this analysis are reported hereafter:  

\begin{itemize}
    \item Wider bandwidths and/or subcarrier spacings lead to lower \acp{RTT}, whereas employing higher modulation orders is beneficial only when the data plane resources saturate;
    \item \ac{1D-CNN} is the best-performing \ac{DL} model for \ac{RUL} estimation as it shows the best trade-off between cost and average advance for all the considered margins;
    \item The use of dedicated \ac{RAN} and \ac{5CN} resources, as in Architecture 3 and 4, together with their network settings result in a reduction of the average \ac{RTT} w.r.t other solutions, like Architecture 2 and 1;
    \item The training of \ac{1D-CNN} for \ac{RUL} estimation needs a margin $m \geq 10$ to provide an average advance time compatible with the average \ac{RTT} performance provided by all the considered network architectures;
    \item The training of \ac{1D-CNN} for \ac{RUL} estimation can leverage a margin $m = 5$, which features lower costs, for certain architectures and number of \acp{AGV} $N$. However, if the \ac{RTT} increases, due to an increase of $N$ or to a less performing architecture, using the set up with $m = 10$ is required;
    \item The average \ac{RTT} is mainly affected by application-specific parameters, i.e., $T_{\rm P\_S}$ and $T_{\rm A}$, rather than \ac{5G NR}-related optimizations.
\end{itemize}
\vspace{-0.5mm}
In future works, we aim to address more realistic, and therefore complex, scenarios by considering multiple \acp{gNB} and different channel models. Our future research will also target reliability assessment for safety critical \ac{RUL}-based chains.

\section*{Acknowledgement}

This work was partially supported by the European Union under the Italian National Recovery and Resilience Plan (NRRP) of NextGenerationEU, partnership on “Telecommunications of the Future” (PE00000001 - program “RESTART”).




\bibliographystyle{ieeetr}
\bibliography{references.bib}

\end{document}